\documentclass[letterpaper, 10 pt, conference]{ieeeconf}
\IEEEoverridecommandlockouts
\usepackage{amsmath, amssymb, stmaryrd, dsfont, mathtools}
\usepackage{booktabs}
\usepackage{subcaption}
\usepackage{tikz, pgfplots}
\usepackage{etoolbox}
\pgfplotsset{compat=newest}
\usetikzlibrary{arrows, positioning, calc}

\newlength\figureheight
\newlength\figurewidth

\newtheorem{theorem}{Theorem}
\newtheorem{proposition}{Proposition}

\newtheorem{problem}{Problem}
\newtheorem{lemma}{Lemma}
\newtheorem{remark}{Remark}

\newcounter{mytempeqncnt}

\newcommand{\blkdiag}[1]{\text{blkdiag}\left( #1 \right)}
\newcommand{\fullmat}[1]{\text{fullmat}\left( #1 \right)}

\title{\LARGE \bf Synthesis of separable controlled invariant sets for modular local control design}

\author{Petter Nilsson and Necmiye Ozay
\thanks{Dept. of
        Electrical Engineering and Computer Science, University of Michigan, Ann Arbor, MI.
        {\tt\small {\{pettni,necmiye\}@umich.edu}}. 
}
}

\begin{document}

\maketitle
\thispagestyle{empty}
\pagestyle{empty}

\begin{abstract} Many correct-by-construction control synthesis methods suffer from the curse of dimensionality. Motivated by this challenge, we seek to reduce a correct-by-construction control synthesis problem to subproblems of more modest dimension. As a step towards this goal, in this paper we consider the problem of synthesizing decoupled robustly controlled invariant sets for dynamically coupled linear subsystems with state and input constraints. Our approach, which gives sufficient conditions for decoupled invariance, is based on optimization over linear matrix inequalities which are obtained using slack variable identities. We illustrate the applicability of our method on several examples, including one where we solve local control synthesis problems in a compositional manner.
\end{abstract}

\section{Introduction} 
\label{sec:introduction}

Distributed embedded control systems are already integral parts of many safety-critical systems, including those in avionics \cite{tallantVV,ozay2011distributed}, automotive \cite{autodes}, electricity generation and distribution \cite{massoud_05_toward}, and medical equipment \cite{healthcps}. Due to increasing complexity and integration of a large number of components and subsystems, certification of safety and performance properties of such systems constitutes a bottleneck in terms of design time. This burden can be partially alleviated by adopting formal methods-based verification and correct-by-construction control synthesis techniques \cite{tabuada2009verification}.

A fundamental property related to safety is invariance \cite{blanchini1999set}. In this paper, we consider the problem of synthesizing polytopic controlled invariant sets for distributed systems consisting of a set of linear constrained subsystems. We assume that each subsystem has its own controller that has some local sensing capabilities and that is required to achieve a local safety specification, while there is coupling between subsystems through the dynamics. Our goal is to synthesize a separable controlled invariant set, which essentially is a cross-product of local invariant sets for the subsystems. The behaviors of the subsystems can then be decoupled as long as their states are constrained to these sets. As such, it is possible to compositionally synthesize more advanced controllers (for instance, from temporal logic specifications, or using model predictive control) within these sets.

Compositional approaches have attracted considerable attention in recent years in the context of verification of stability \cite{topcu2009compositional}, safety \cite{sloth2012compositional}, and performance \cite{meissen2014performance} specifications of dynamical systems. The results on compositional synthesis are mostly limited to linear systems. Our work also falls into this category and is tightly related to the compositional synthesis approaches proposed for finding ellipsoidal controlled invariant sets for linear systems to be used as terminal sets in model predictive control  \cite{jokic2009decentralized,rakovic2010practical,conte2012distributed,giselsson2014onfeasibility}. Instead of searching for ellipsoidal invariant sets, we search for polytopic invariant sets with tunable complexity. Polytopic sets are more commonly used in temporal logic-based control synthesis techniques \cite{kloetzer2008fully,wongpiromsarn2012receding,gol2014finite, nilsson2014preliminary}. Therefore, our approach constitutes a first step in extending some of these techniques to the distributed setting. 

Our main contribution is to demonstrate how local invariant sets can be used to separately synthesize local controllers to achieve local tasks while still guaranteeing correctness when these controllers are composed together. This is achieved by (i) synthesizing feedback controllers that render a subset of the state-space of each subsystem robustly invariant, (ii) extracting the set of allowable inputs that will keep these sets invariant even when the feedback controllers are discarded, and (iii) solving a local synthesis problem with complex task specifications, for instance, expressed in temporal logics, with the new state and input constraints. For the first step we build on recent results on low complexity polytopic invariant set computation. In particular, we extend the results in \cite{Tahir:2014cz} from the centralized setting to the decentralized setting and allow synthesis of local invariant sets in the form of arbitrary zonotopes, where the number of generators is an input to our method, as opposed to linearly transformed hyper boxes as in \cite{Tahir:2014cz}. Although conceptually simple, the computation of these local invariant sets is far from trivial, as it requires solving a non-convex feasibility problem. In order to overcome this difficulty, we resort to slack variable based relaxations and provide linear matrix inequality (LMI)-based sufficient conditions for computing these invariant sets.

We have structured this paper in the following way. After giving preliminaries on notation and three key results from \cite{Tahir:2014cz} in Section \ref{sec:preliminaries}, we formally introduce the decoupled invariance problem in Section \ref{sec:problem_formulation}. Next, we present the main results of this paper in Section \ref{sec:main_res} and discuss where conservativeness is introduced. We give some examples in Section \ref{sec:examples} to illustrate the advantages of our method, before concluding the paper in Section \ref{sec:conclusions}.

\section{Preliminaries} 
\label{sec:preliminaries}

In this paper we will make use of the following notation; $I_n$ denotes the identity matrix of size $n \times n$. When the dimension is apparent from the context, the subscript will be dropped. We denote by $e_i$ the $i$:th Euclidean standard basis column vector, while $\mathds{1}$ is a column vector where all entries are $1$. Furthermore, given a one-dimensional set of matrices $\{ A_i \}_i = \{A_1, A_2 , \ldots \}$, we denote by $\blkdiag{ \{ A_i \}_i }$ the block-diagonal matrix formed by that set. Similarly, for a two-dimensional set of matrices $\{ A_{ij} \}_{ij}$, we write $\fullmat{ \{ A_{ij} \}_{ij} }$ to indicate the block matrix whose $(i,j)$'th block is $A_{ij}$. When we write $\Pi_i \mathcal X_i$ we mean the cross product between sets, i.e. $\Pi_{i=1}^2 \mathcal X_i = X_1 \times X_2 = \{ (x_1, x_2) : x_1 \in \mathcal X_1, x_2 \in \mathcal X_2 \}$. For two sets $\mathcal X$ and $\mathcal Y$, we denote the Minkowski sum by $\mathcal X \oplus \mathcal Y = \{ x + y : x \in \mathcal X, y \in \mathcal Y \}$.

When we introduce matrix variables, we will use ``(full)'', ``(sym)'',  or ``(diag)'' to indicate if this matrix variable is a full matrix, or restricted to being symmetric or diagonal. In the case of symmetric matrices, we will write $*$ for entries whose values follow from symmetry.

We now present three lemmas that are crucial to the developments in later sections. First, we recall the following result from \cite{Tahir:2014cz} which we state without proof.
\begin{lemma}
\label{lemma:thlemma}
	Let $R$ (sym), $Z$ (sym), $A$ (full) and $B$ (full) be arbitrary matrices. Then the following conditions are equivalent
	\begin{enumerate}
		\item
	\begin{equation}
	\label{eq:ABlemma1}
		\begin{bmatrix}
			R & AB \\ * & Z
		\end{bmatrix} \succ 0.
	\end{equation}
	\item
	\begin{equation}
		\exists X \; \text{(sym)} : \begin{bmatrix}
			R & A \\ * & X^{-1}
		\end{bmatrix} \succ 0, \; \begin{bmatrix}
			X & B \\ * & Z
		\end{bmatrix} \succ 0
	\end{equation}
	\end{enumerate}
\end{lemma}
The usefulness of this lemma is to separate the matrix product $AB$ in the upper right entry of \eqref{eq:ABlemma1}. However, the introduced matrix variable $X$ appears both as itself and its inverse.

The next two lemmas are inspired by the same paper, but we have altered the formulation to our purposes. For completeness, we provide proofs in the Appendix.

\begin{lemma}
\label{corr:result1b}
The following two statements are equivalent
\begin{enumerate}
	\item
	\begin{equation}
		\begin{bmatrix}
			C^T X C & Y  \\
			* &  Z
		\end{bmatrix} \succ 0
	\end{equation}		
	\item $C$ is non-singular, $X \succ 0$, and there exists $\Psi$ such that
	\begin{equation}
		\begin{bmatrix}
			C \Psi + \Psi^T C^T - X^{-1} & \Psi^T Y \\ * & Z
		\end{bmatrix} \succ 0.
	\end{equation}
\end{enumerate}
\end{lemma}

\begin{lemma}
\label{lemma:result2}
	If there exist $\Theta$ (full), $\Gamma$ (sym) and $\Xi$ (sym) such that
	\begin{equation}
	\label{eq:lemma2cond1}
	\Delta \dot = \left[\begin{smallmatrix}
		\Gamma & Y \\ * & \Xi
	\end{smallmatrix}\right] \succ 0, \quad
		\left[\begin{smallmatrix}
			Z + \Xi & \left[\begin{smallmatrix}
				-X & I
			\end{smallmatrix}\right] \Theta & V \\
			* & \Theta + \Theta^T - \Delta & 0 \\ * & * & W
		\end{smallmatrix}\right] \succ 0,
	\end{equation}
	then \begin{equation}
		\left[\begin{smallmatrix}
			 Z + X Y + Y^T X^T & V \\ * & W
		\end{smallmatrix}\right] \succ 0.
	\end{equation}
\end{lemma}

\begin{remark}
	As is evident from the proof, if $X \Gamma X^T$ is added to the top left entry of the right hand matrix in \eqref{eq:lemma2cond1}, the reverse implication also holds.
\end{remark}

\section{Problem formulation} 
\label{sec:problem_formulation}

As alluded to in Section \ref{sec:introduction}, as a first step in modular local control design, we seek to find a separable invariant set for a system, where the splitting into subsystems is given beforehand. Specifically, we consider a discrete time system of the form 
\begin{equation}
\label{eq:systemmodel}
	x_i(t+1) = A_{ii} x_i(t) + \sum_{j \neq i} A_{ij} x_j(t) + B_i u_i(t) + E_i d_i(t),
\end{equation}
for $i = 1, \ldots d$, where $x_i \in \mathbb{R}^{n_i}$ is the state of subsystem $i$. As can be seen, each subsystem is associated with an input $u_i \in \mathbb{R}^{m_i}$ and a disturbance $d_i \in \mathbb{R}^{p_i}$, both of which do not affect other subsystems. We assume that the inputs of \eqref{eq:systemmodel} are bounded as
\begin{equation}
	\label{eq:inputbounds}
	u_i(t) \in \mathcal U_i \; \dot = \; \{ u_i :  H_u^i u_i \leq h_u^i \}, \quad H_u^i \in \mathbb{R}^{\mathcal N_u^i \times m_i},
\end{equation}
for all $i$, and that there are bounds on disturbance given by
\begin{equation}
	\label{eq:distass}
	d_i(t) \in \mathcal D_i \; \dot = \; \{ d_i : -\mathds{1} \leq H_d^i d_i \leq \mathds{1} \},
\end{equation}
where $H_d^i \in \mathbb{R}^{p_i \times p_i}$ is a non-singular square matrix. We also consider state constraints of the form
\begin{equation}
	\label{eq:stateconstr}
	x_i \in \mathcal S_i \; \dot = \; \{ x_i : H_s^i x_i \leq h_s^i \}, \quad H_s^i \in \mathbb{R}^{\mathcal N_s^i \times n_i}.
\end{equation}
We are now ready to state the first problem we seek to solve.
\begin{problem}
\label{probl:main}
Given a set of dynamically coupled linear subsystems \eqref{eq:systemmodel} together with input constraints, disturbance bounds and state constraints of the forms \eqref{eq:inputbounds} - \eqref{eq:stateconstr}, find for all $i = 1, \ldots, d$ sets $\mathcal X_i \subset \mathcal S_i$ such that
	\begin{equation}
	\begin{aligned}
		\label{eq:invariancespec}
			\forall x_i(t) \in \mathcal X_i \; \; \exists u_i(t)  \in \mathcal U_i  \; \; \forall x_j(t) \in \mathcal X_j \; \; \forall d_i(t) \in \mathcal D_i, \\
			x_i(t+1) \in \mathcal X_i.
	\end{aligned}
	\end{equation}
\end{problem}

This problem exhibits some interesting features due to its circular nature. A solution is basically an assume-guarantee protocol \cite{frehse2004assume} where each subsystem guarantees that it will limit its effect on other subsystems, under the assumption that the other subsystems do the same. Due to this nature, it is not trivial to find invariant sets using classical iterative methods such as \cite{Bertsekas:1972gm,DeSantis:2004et}. The advantage of separable invariant sets, once obtained, is that a controller for subsystem $i$ will not affect the safety of the other subsystems adversely as long as $x_i$ remains in $\mathcal X_i$. This allows for local control objectives to be pursued as long as the constraints $x_i \in \mathcal X_i$ are complied with.

Here we outline two possible modifications to the problem statement, which may be useful in applications. Firstly, the formulation can be modified to reflect the information availability for local controllers. While in Problem \ref{probl:main} the local controllers have access only to their own state, there may be situations where the state of some of the other subsystems can be measured locally. In that case, this information can be leveraged when computing a local control input. Formally, such information availability can be incorporated by swapping quantifiers for states and input in \eqref{eq:invariancespec}, and it can be handled in our solution framework as explained later in Remark \ref{remark:selectK}. Secondly, it may be desirable to impose additional constraints on the local invariant sets $\mathcal X_i$. For instance, if there are certain subsets $ {\mathcal Y}_{i,j}$ for some $j=1,\ldots, n_{i,J}$ of the subsystem $i$ that are known to be important for satisfaction of local control objectives, it is possible to impose set containment constraints of the form $\mathcal X_i \supset \cup_{j=1}^{n_{i,J}}\mathcal Y_{i,j}$ in our framework, as briefly discussed in Remark \ref{remark:containment}.

\section{Main results} 
\label{sec:main_res}
\begin{figure*}[!t]
\normalsize
\setcounter{mytempeqncnt}{\value{equation}}
\setcounter{equation}{12}
\vspace*{3pt}
\begin{equation}
	\label{eq:lmi_first}
	 \left[\begin{smallmatrix}
		\Gamma_j & \Psi_j \\ * & \Xi_j
		\end{smallmatrix}\right] \succ 0, \quad
	\left[\begin{smallmatrix}
		\Xi_j - \Phi_j^{-1} & \Omega^1_j- \left[\begin{smallmatrix}
		H_x^{-1} & 0 \\ 0 & H_x^{-1}
		\end{smallmatrix}\right] & \Omega^2_j  -  \left[\begin{smallmatrix}
		H_x^{-1} & 0 \\ 0 & H_x^{-1}
		\end{smallmatrix}\right]  & \Psi_j^T \left[\begin{smallmatrix}
			Z^T e_j  \\ Z^T e_j 
		\end{smallmatrix}\right] \\
		* & 2 \left[\begin{smallmatrix}
		\Lambda & 0 \\ 0 & \Lambda
		\end{smallmatrix}\right] - \Gamma_j & \left[\begin{smallmatrix}
		\Lambda & 0 \\ 0 & \Lambda
		\end{smallmatrix}\right] + (\Omega^1_j)^T - \Psi_j & 0 \\
		* & * & \Omega^2_j + (\Omega^2_j)^T - \Xi_j & 0 \\
		* & * & * & \lambda_{i(j)} - \mathds{1}^T D_x^j \mathds{1}  - \mathds{1}^T D_d^j \mathds{1}
	\end{smallmatrix}\right] \succ 0,\\	
\end{equation}
\setcounter{equation}{\value{mytempeqncnt}}

\hrulefill
\vspace*{-8pt}
\end{figure*}

\subsection{Computation of separable invariant sets}
\label{sub:comp_sep_inv}
In order to attack Problem \ref{probl:main}, we restrict the description in two ways. Firstly, we do not consider arbitrary nonlinear controllers but search for sets that can be rendered invariant by local feedback controllers $u_i = K_i x_i$. Secondly, we restrict the sets $\mathcal X_i$ to be symmetric zonotopes with a fixed number of generators. Specifically, we search for invariant sets $\mathcal X_i$ of the form
\begin{equation}
	\mathcal X_i = \left\{ x_i : -\mathds{1} \leq Z_i H_x^i x_i \leq \mathds{1} \right\},
\end{equation}
where $Z_i \in \mathbb{R}^{\mathcal N_x^i \times n_i}$ is an arbitrary given matrix and $H_x^i \in \mathbb{R}^{n_i \times n_i}$ is non-singular. $\mathcal N_x^i$ gives the number of generators of the zonotope describing the set $\mathcal X_i$, whereas $H_x^i$ is a variable that represents an \emph{unknown} linear transformation. This description allows us to tune the complexity of the invariant sets we search for. 

Before stating the main result, we concatenate the indexed variables introduced above to obtain a compact representation of the overall system. Specifically, we let $A = \fullmat{\{A_{ij}\}_{ij}}$, $B = \blkdiag{\{B_i\}_i}$, $K = \blkdiag{\{K_i\}_i} $, $ E = \blkdiag{ \{ E_i \}_i }$, $Z = \blkdiag{ \{Z_i\}_i }$, $H_x = \blkdiag{ \{H_x^i\}_i }$, $H_u = \blkdiag{ \{H_u^i\}_i }$, $h_u = [(h_u^1)^T, \ldots, (h_u^d)^T]^T$, $H_d = \blkdiag{ \{H_d^i\}_i }$, $H_s = \blkdiag{ \{H_s^i\}_i }$, $h_s = [(h_s^1)^T, \ldots, (h_s^d)^T]^T$, $x = \left[\begin{smallmatrix} x_1^T & \cdots & x_d^T \end{smallmatrix}\right]^T \in \mathbb{R}^n$, $u = \left[\begin{smallmatrix} u_1^T & \cdots & u_d^T \end{smallmatrix}\right]^T \in \mathbb{R}^m$ and $d = \left[\begin{smallmatrix} d_1^T & \cdots & d_d^T \end{smallmatrix}\right]^T \in \mathbb{R}^p$. The dimensions of these composed variables become $n = \sum_i n_i$ and similarly for $m$ and $p$. The number of rows of $Z$, $H_u$, and $H_s$ are $\mathcal N_x = \sum_i \mathcal N_x^i$, $\mathcal N_u= \sum_i \mathcal N_u^i$ and $\mathcal N_s = \sum_i \mathcal N_s^i$, respectively. For $A_K \dot = A+BK$, the closed loop dynamics then become
\addtocounter{equation}{1}
\begin{equation}
	x(t+1) = A_K x(t) + E d(t).
\end{equation}
The control and state constraints are given by $\mathcal X = \Pi_i \mathcal X_i$ and $\mathcal D = \Pi_i \mathcal D_i$, which can be compactly represented as
\begin{equation}
\begin{aligned}
	\mathcal X = \{ x : -\mathds{1} \leq Z H_x x \leq \mathds{1} \}, \\
	\mathcal D = \{ d : -\mathds{1} \leq H_d d \leq \mathds{1} \}.
\end{aligned}
\end{equation}
The conditions in Problem \ref{probl:main} can now be stated as follows using set-theoretic constructs.
\begin{enumerate}
	\item Separable invariance of closed loop system: 
	\begin{equation}
		\label{eq:invariance}
		A_K  \mathcal X \oplus E \mathcal D \subset \mathcal X.
	\end{equation}
	\item State constraints: 
	\begin{equation}
		\label{eq:stateconst}
		\mathcal X \subset \Pi_i \mathcal S_i = \{ x : H_s x \leq h_s \}.
	\end{equation}
	\item Control constraints: 
	\begin{equation}
		\label{eq:contconstr}
		BK \mathcal X \subset \Pi_i \mathcal U_i = \{ u : H_u u \leq h_u \}.
	\end{equation}
\end{enumerate}

\begin{theorem}
\label{thm:main}
	If there exists $\Lambda \dot = \blkdiag{ \{ I_{n_i} \lambda_i \}_{i=1}^d } \succ 0$, and for all $j=1, \ldots, \mathcal N_x$ there exist matrix variables $D_x^j \succ 0$ (diag), $\Phi_j^{-1}$ (sym), $\Gamma_j$ (sym), $\Xi_j$ (sym), $\Psi_j$ (full), $\Omega^1_j$ (full), $\Omega^2_j$ (full), and for all $k = 1, \ldots \mathcal N_s$ matrix variables $D_s^k \succ 0$ (diag), and for all $l = 1, \ldots \mathcal N_u$ matrix variables $D_u^l \succ 0$ (diag), such that LMI's \eqref{eq:lmi_first} and \eqref{eq:lmi_third} - \eqref{eq:lmi_fifth} are satisfied, then the block diagonal pair $(H_x, \hat K H_x)$ constitutes a solution to Problem \ref{probl:main} when appropriately decomposed.
\begin{align}
\label{eq:lmi_third}
& \left[\begin{smallmatrix}
		\begin{smallmatrix}
			Z^T  D_x^j Z & 0 \\ *  &  D_d^j
		\end{smallmatrix} & \begin{smallmatrix}
			-\frac{1}{2} (H_x^{-T} A^T + \hat K^T B^T) & 0 \\ 
			0 & -\frac{1}{2} H_d^{-T} E^T 
		\end{smallmatrix} \\
		* & \left[\Phi_j^{-1} \right]
	\end{smallmatrix}\right] \succ 0 \\
\label{eq:lmi_fourth} & \left[\begin{smallmatrix}
		Z^T D_s^k Z & - \frac{1}{2} H_x^{-T} H_s^T e_k \\ * & e_k^T h_s - \mathds{1}^T D_s^k \mathds{1}
	\end{smallmatrix}\right] \succ 0, \\
\label{eq:lmi_fifth}
	& \left[\begin{smallmatrix}
		Z^T D_u^l Z& - \frac{1}{2} \hat K^T H_u^T e_l \\ * & e_l^T h_u - \mathds{1}^T D_u^l \mathds{1}
	\end{smallmatrix}\right] \succ 0
\end{align}
\end{theorem}
\vspace{2mm}
\begin{proof} Given in the appendix.
\end{proof}

\begin{remark}
	In \eqref{eq:lmi_first}, $i(j)$ is used to denote the smallest index $i$ s.t. $j \leq \sum_{k=1}^i \mathcal N_x^k$.
\end{remark}

\vspace{1mm}

\begin{remark}
\label{remark:selectK}
	The feedback matrix $K$ was defined above to be block diagonal, to only allow local state information to influence the control signal computation. If more information is available locally (i.e. the states of neighboring subsystems are available to the controller), this restriction can be relaxed by allowing additional non-zero blocks in $K$. We illustrate this with an example in Section \ref{sub:find_inv_sets} where neighboring subsystems are allowed to interchange state information. 
\end{remark}

\begin{remark}
\label{remark:containment}	
	Set containment constraints of the type $\mathcal X_i \supset \cup_{j=1}^{n_{i,j}} \mathcal Y_{i,j}$ can be handled similarly to the invariance constraint \eqref{eq:invariance}, at the cost of some additional conservatism. This gives rise to three additional LMI's on the same forms as those in \eqref{eq:lmi_first} and \eqref{eq:lmi_fourth}, but the details are omitted in this paper.
\end{remark}

\begin{remark}
\label{remark:chooseZ}
	A natural way to select $Z_i$ is to pick (randomly or evenly spaced) unit vectors from $S^{n_i-1}_+ \dot = \{ (g_1, \ldots g_{n_i}) \in S^{n_i-1} : g_1 \geq 0 \}$, where $S^n$ is the $n$-dimensional unit sphere.
\end{remark}

\begin{remark}
	The sizes of the matrices in \eqref{eq:lmi_first} and \eqref{eq:lmi_third} - \eqref{eq:lmi_fifth} are $4n \times 4n$, $6n+1 \times 6n+1$, $4n \times 4n$, $n+1 \times n+1$ and $n+1 \times n+1$, respectively. When there is no external disturbance, there is no need to introduce the variables $D_d^i$ and the sizes of the first three matrices reduce to $2n \times 2n$, $3n+1 \times 3n+1$ and $2n \times 2n$, which is a computationally easier problem.
\end{remark}

\subsection{Using separable invariant sets for compositional synthesis} 
\label{sub:using_separable_invariant_sets_for_compositional_synthesis}
In general, $K_i x_i$ is not the unique control signal that renders $\mathcal X_i$ invariant. By performing polytopic reachability computations, the envelope of invariant-enforcing controls can be extracted for a given point. For such a point $x_i^0 \in \mathcal X_i$, the set of all inputs that both satisfy the local control constraints and guarantees invariance of $\mathcal X_i$ is the set of $u_i$'s that satisfy
\begin{equation}\label{eq:newinputconst}
	\left[\begin{smallmatrix}
	 	H_u^i \\
	 	Z_i H_x^i B  \\
	 	- Z_i H_x^i B 
	\end{smallmatrix} \right] { u_i }\leq \left[\begin{smallmatrix}
	   h_u^i \\
	   \mathds{1} - H_x^i A x_i^0 - \sum\limits_{j \neq i} \max\limits_{x_j \in \mathcal X_j} H_x^i A_{ij} x_j \\ 
	   - \mathds{1} + H_x^i A x_i^0 + \sum\limits_{j \neq i}\min\limits_{x_j \in \mathcal X_j} H_x^i A_{ij} x_j
	\end{smallmatrix} \right],
\end{equation}
where the inequality should hold element-wise. By construction this set will be non-empty as long as $x_i^0 \in \mathcal X_i$. The $\max$ and $\min$ terms in this expression can be found by solving linear programs over the other invariant sets $\mathcal X_j$, or by enumerating their vertices. This flexibility in control signal selection can be exploited to design local controllers separately in a compositional manner, for instance with the goal of performing local control tasks. In what follows we assume that the local task specifications are given in terms of linear temporal logic (LTL) over atomic propositions defined on the state-spaces of individual subsystems.{\footnote{Note that the choice of LTL for specifying local tasks is arbitrary; other specification languages can be used as well. Therefore, we skip the details of LTL and refer the interested readers to \cite{Baier:2008vh}.}} To summarize, by construction we have the following result which enables local controller synthesis with global guarantees. 

\begin{proposition}
For a given system in the form \eqref{eq:systemmodel} - \eqref{eq:stateconstr}, let $\mathcal X_i$ be invariant sets that satisfy the requirements of Problem \ref{probl:main}, and let $\varphi_i$ be a local LTL specification for subsystem $i$ for $i = 1, \ldots, d$. Furthermore, for $i = 1, \ldots, d$, let $u_i : \mathcal X_i \rightarrow \mathcal U_i$ be local controllers that generate closed loop trajectories that satisfy $\varphi_i$ while also satisfying the (state-dependent) input constraint \eqref{eq:newinputconst}. Then, the composed control $u \doteq [u_1, \ldots u_{d}] : \Pi_i \mathcal X_i \rightarrow \Pi_i \mathcal U_i$ generates closed loop behaviors of the overall system that satisfy $\land_i \varphi_i$.
\end{proposition}


\subsection{Discussion about conservativeness} 
\label{sub:discussion_about_conservativeness}

For the separable invariant set computation, conservativeness enters the proof of Theorem \ref{thm:main} at two places. First, a positive term $\bar \Lambda^{-1} \bar H_x^{-T} \Gamma_j \bar H_x^{-1} \Lambda^{-1}$ is thrown away when Lemma \ref{lemma:result2} is used on \eqref{eq:pre_result1b}. Secondly, instead of allowing an arbitrary $\Theta_j$ in \eqref{eq:post_result1b}, the upper two blocks of $\Theta_j$ is restricted to be equal to $\Lambda$ to make the resulting inequality linear. While it is easy ex post to evaluate the effect of ignoring the positive term by looking at its magnitude, it is more subtle how much conservativeness that is introduced by restricting $\Theta_j$. Once an initial solution is obtained, it can be iteratively updated as in \cite{Tahir:2014cz} by using the values of LMI variables  $\Phi_j^0$ and $H_x^0$ from a previous iteration. The idea is to use a different matrix inverse identity than the one employed in the proof Lemma \ref{lemma:result2}. The result is that the only term that needs to be discarded, and thus introduces conservativeness, is a quadratic term that is small if $\Phi_j^0$ and $\Phi_j$, and $\bar H_x$ and $\bar H_x^0$, are close. Therefore a solution can be iteratively updated without much conservativeness by performing small steps.

For the overall synthesis problem, first computing the local invariant sets and then trying to solve local synthesis problems in each invariant set is clearly more conservative than trying to synthesize local controllers by taking into account all the interactions. However, there are trade-offs between conservativeness, modularity of the local control design and computational complexity. Correct-by-construction control synthesis from temporal logic specifications is computationally challenging for high dimensional systems in general \cite{tabuada2009verification,wongpiromsarn2012receding}. Therefore, decomposing the problem into smaller subproblems per subsystem improves scalability \cite{ozay2011distributed}. Moreover, the local robust controlled invariant sets provide a modular framework in that it is possible to replace an existing local controller for a subsystem, for instance in order to pursue a different local control objective, without needing to resynthesize the rest of the local controllers.


\section{Examples} 
\label{sec:examples}

We illustrate the applicability of this approach on a few examples. First, in Section \ref{sub:find_inv_sets} we compute separable invariant sets for two different systems. Thanks to the increased flexibility in our set description, we are able to solve problems where the geometry is not compatible with linearly transformed hyper boxes. Then, in Section \ref{sub:robot_uav}, we look at a system of connected mobile robots and show how our framework allows local control tasks to be performed while guaranteeing overall safety. 

\subsection{Finding invariant sets} 
\label{sub:find_inv_sets}

The dynamics of our first example are as follows. For each subsystem $x_i \in \mathbb{R}^2$, the evolution consists of a rotational part and disturbance coming both from the other subsystems and from an additive term. 
\begin{equation}
	x_i(t+1) = \alpha_i R( \theta ) x_i(t) + u_i(t) + \sum_{i \neq j} \beta_{ij} x_j + d_i(t).
\end{equation}
Here $R(\theta) \in SO(2)$ is the (counter-clockwise) rotation matrix. The input $u_i \in \mathbb{R}^2$ and disturbance $d_i \in \mathbb{R}^2$ are bounded by $\| u_i \|_\infty \leq u_{max, \; i}$ and $\| d_i \|_\infty \leq d_{max, \; i}$.

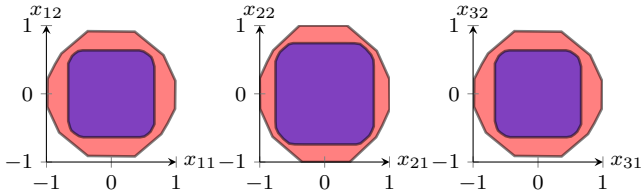
\begin{figure}[tb]
	\begin{center}
		\begin{subfigure}{0.3\columnwidth}
			\setlength\figurewidth{0.7\columnwidth} 
			\setlength\figureheight{0.7\columnwidth} 
			\footnotesize 
%
\begin{tikzpicture}

\begin{axis}[%
width=0.95092\figurewidth,
height=\figureheight,
at={(0\figurewidth,0\figureheight)},
scale only axis,
every outer x axis line/.append style={black},
every x tick label/.append style={font=\color{black}},
xmin=-1,
xmax=1,
xmajorgrids,
every outer y axis line/.append style={black},
every y tick label/.append style={font=\color{black}},
ymin=-1,
ymax=1,
ymajorgrids,
every outer z axis line/.append style={black},
every z tick label/.append style={font=\color{black}},
zmin=-1,
zmax=1,
zmajorgrids,
view={0}{90},
xlabel={$x_{11}$},
ylabel={$x_{12}$},
axis x line*=bottom,
axis y line*=left,
axis z line*=left,
xmajorgrids=false, ymajorgrids=false,
xmajorgrids=false, ymajorgrids=false, axis x line=bottom, axis y line=left, every axis x label/.style={at={(current axis.south east)},anchor=west},  every axis y label/.style={at={(current axis.north west)},anchor=south}
]

\addplot3[area legend,solid,line width=1.0pt,fill=red,opacity=5.000000e-01,draw=black,forget plot]
table[row sep=crcr] {%
x	y	z\\
0.981947695834529	-0.219999663983693	0\\
0.8677351339421	-0.433086520589545	0\\
0.770344551796216	-0.589951864378463	0\\
0.363890187342165	-0.918076291026008	0\\
-0.361580972202966	-0.910810523559532	0\\
-0.803798566218363	-0.566962241827805	0\\
-0.897077269672788	-0.375630860353224	0\\
-0.986386842662562	-0.191149565115055	0\\
-0.981947695834529	0.219999663983767	0\\
-0.867735133942711	0.433086520588562	0\\
-0.770344551796216	0.589951864378462	0\\
-0.363890187342165	0.918076291026007	0\\
0.361580972202966	0.910810523559532	0\\
0.803798566218363	0.566962241827805	0\\
0.897077269672799	0.375630860353202	0\\
0.986386842662562	0.191149565115055	0\\
}--cycle;

\addplot3[area legend,solid,line width=1.0pt,fill=blue,opacity=5.000000e-01,draw=black,forget plot]
table[row sep=crcr] {%
x	y	z\\
0.663799942055636	-0.444088649288143	0\\
0.652666438430612	-0.467515533399965	0\\
0.64312120398826	-0.484775970313249	0\\
0.604955220583771	-0.553680354206922	0\\
0.564440977760244	-0.590007008634071	0\\
0.536952227283462	-0.608053027987019	0\\
0.510421875911391	-0.625369336427388	0\\
0.442968529076963	-0.63693309513251	0\\
-0.357031470923037	-0.63693309513251	0\\
-0.430575000089055	-0.636932915190674	0\\
-0.470113187550046	-0.628060674243044	0\\
-0.499764184883229	-0.619761032729448	0\\
-0.570082209164957	-0.575470387224493	0\\
-0.615615781359923	-0.538353529396928	0\\
-0.625466922610587	-0.517532879597072	0\\
-0.634901564981901	-0.497456965741523	0\\
-0.663396551449252	-0.400876882122716	0\\
-0.663799942055633	-0.355911350711853	0\\
-0.663799942055633	0.444088649288147	0\\
-0.65266643843059	0.467515533400004	0\\
-0.64312120398826	0.484775970313249	0\\
-0.604955220583771	0.553680354206922	0\\
-0.564440977760244	0.590007008634071	0\\
-0.536952227283465	0.608053027987017	0\\
-0.510421875911391	0.625369336427388	0\\
-0.442968529076951	0.636933095132512	0\\
0.357031470923049	0.636933095132512	0\\
0.430575000089068	0.636932915190676	0\\
0.47011318754986	0.628060674243096	0\\
0.499764184883229	0.619761032729448	0\\
0.570082209164957	0.575470387224493	0\\
0.615615781359923	0.538353529396928	0\\
0.625466922610583	0.51753287959708	0\\
0.634901564981901	0.497456965741523	0\\
0.663396551449252	0.400876882122716	0\\
0.663799942055636	0.355911350711857	0\\
}--cycle;

\end{axis}
\end{tikzpicture}%
		\end{subfigure}~ 
		\begin{subfigure}{0.3\columnwidth}
			\setlength\figurewidth{0.7\columnwidth} 
			\setlength\figureheight{0.7\columnwidth} 
			\footnotesize 
%
\begin{tikzpicture}

\begin{axis}[%
width=0.95092\figurewidth,
height=\figureheight,
at={(0\figurewidth,0\figureheight)},
scale only axis,
every outer x axis line/.append style={black},
every x tick label/.append style={font=\color{black}},
xmin=-1,
xmax=1,
xmajorgrids,
every outer y axis line/.append style={black},
every y tick label/.append style={font=\color{black}},
ymin=-1,
ymax=1,
ymajorgrids,
every outer z axis line/.append style={black},
every z tick label/.append style={font=\color{black}},
zmin=-1,
zmax=1,
zmajorgrids,
view={0}{90},
xlabel={$x_{21}$},
ylabel={$x_{22}$},
axis x line*=bottom,
axis y line*=left,
axis z line*=left,
xmajorgrids=false, ymajorgrids=false,
xmajorgrids=false, ymajorgrids=false, axis x line=bottom, axis y line=left, every axis x label/.style={at={(current axis.south east)},anchor=west},  every axis y label/.style={at={(current axis.north west)},anchor=south}
]

\addplot3[area legend,solid,line width=1.0pt,fill=red,opacity=5.000000e-01,draw=black,forget plot]
table[row sep=crcr] {%
x	y	z\\
0.999796282401273	-0.229825786917068	0\\
0.888461246151561	-0.464094628034253	0\\
0.793008901727485	-0.636698997168105	0\\
0.387866473492215	-0.999965541439603	0\\
-0.34756881816797	-0.999963742021243	0\\
-0.802904540117625	-0.628795163745592	0\\
-0.901415952624249	-0.42058866574706	0\\
-0.995762376337405	-0.219829527191545	0\\
-0.999796282401212	0.229825786917087	0\\
-0.888461246150788	0.464094628035649	0\\
-0.793008901727484	0.636698997168106	0\\
-0.387866473492215	0.999965541439603	0\\
0.34756881816797	0.999963742021243	0\\
0.802904540117624	0.628795163745593	0\\
0.901415952624252	0.420588665747053	0\\
0.995762376337405	0.219829527191545	0\\
}--cycle;

\addplot3[area legend,solid,line width=1.0pt,fill=blue,opacity=5.000000e-01,draw=black,forget plot]
table[row sep=crcr] {%
x	y	z\\
0.757182129069122	-0.393777563348337	0\\
0.756738221889241	-0.434892351328204	0\\
0.756294307206437	-0.476007274238079	0\\
0.742677346173933	-0.508784665096456	0\\
0.729478225732071	-0.540313292794124	0\\
0.718056969542828	-0.561621978454709	0\\
0.706635749351638	-0.582930595694622	0\\
0.696896691137049	-0.598617130073514	0\\
0.687157663242083	-0.614303614263813	0\\
0.635220266796	-0.660245251833432	0\\
0.594574830350595	-0.693057694498187	0\\
0.553929515495314	-0.725870034915415	0\\
0.514625717756477	-0.736301589887105	0\\
0.484218594127241	-0.742389419745668	0\\
-0.315781405872759	-0.742389419745668	0\\
-0.388328313787171	-0.741662855171724	0\\
-0.460875429741684	-0.740936278425076	0\\
-0.546918204310318	-0.72943406713486	0\\
-0.591139842041593	-0.695049357820981	0\\
-0.635361601443133	-0.660664529647808	0\\
-0.714745835752914	-0.573115707270216	0\\
-0.724073682233933	-0.553982633199696	0\\
-0.733401552579375	-0.534849495052238	0\\
-0.742332487048927	-0.516401427302829	0\\
-0.751263444347904	-0.497953297779012	0\\
-0.757182129069122	-0.406222436651663	0\\
-0.757182129069122	0.393777563348337	0\\
-0.75673822188924	0.434892351328206	0\\
-0.756294307206437	0.476007274238089	0\\
-0.742677346173933	0.508784665096464	0\\
-0.729478225732071	0.540313292794133	0\\
-0.718056969542889	0.561621978454612	0\\
-0.706635749351704	0.582930595694516	0\\
-0.696896691137054	0.598617130073506	0\\
-0.687157663242083	0.614303614263813	0\\
-0.635220266795991	0.660245251833426	0\\
-0.594574830350586	0.693057694498181	0\\
-0.553929515495305	0.72587003491541	0\\
-0.51462571775623	0.736301589887155	0\\
-0.48421859412724	0.742389419745668	0\\
0.31578140587276	0.742389419745668	0\\
0.388328313787171	0.741662855171724	0\\
0.460875429741684	0.740936278425076	0\\
0.546918204310318	0.72943406713486	0\\
0.591139842041593	0.695049357820981	0\\
0.635361601443133	0.660664529647808	0\\
0.714745835752914	0.573115707270216	0\\
0.724073682233931	0.5539826331997	0\\
0.733401552579374	0.534849495052239	0\\
0.742332487048928	0.516401427302827	0\\
0.751263444347904	0.497953297779012	0\\
0.757182129069122	0.406222436651663	0\\
}--cycle;

\end{axis}
\end{tikzpicture}%
		\end{subfigure}~ 
		\begin{subfigure}{0.3\columnwidth}
			\setlength\figurewidth{0.7\columnwidth} 
			\setlength\figureheight{0.7\columnwidth} 
			\footnotesize 
%
\begin{tikzpicture}

\begin{axis}[%
width=0.95092\figurewidth,
height=\figureheight,
at={(0\figurewidth,0\figureheight)},
scale only axis,
every outer x axis line/.append style={black},
every x tick label/.append style={font=\color{black}},
xmin=-1,
xmax=1,
xmajorgrids,
every outer y axis line/.append style={black},
every y tick label/.append style={font=\color{black}},
ymin=-1,
ymax=1,
ymajorgrids,
every outer z axis line/.append style={black},
every z tick label/.append style={font=\color{black}},
zmin=-1,
zmax=1,
zmajorgrids,
xlabel={$x_{31}$},
ylabel={$x_{32}$},
view={0}{90},
axis x line*=bottom,
axis y line*=left,
axis z line*=left,
xmajorgrids=false, ymajorgrids=false,
xmajorgrids=false, ymajorgrids=false, axis x line=bottom, axis y line=left, every axis x label/.style={at={(current axis.south east)},anchor=west},  every axis y label/.style={at={(current axis.north west)},anchor=south}
]

\addplot3[area legend,solid,line width=1.0pt,fill=red,opacity=5.000000e-01,draw=black,forget plot]
table[row sep=crcr] {%
x	y	z\\
0.981944911789045	-0.219998809200907	0\\
0.867732709876862	-0.433084981600508	0\\
0.770342430927503	-0.589949823503016	0\\
0.363889282374692	-0.918073227675305	0\\
-0.361579796769419	-0.910807581935862	0\\
-0.803796174082175	-0.566960488797069	0\\
-0.897074638892288	-0.375629748092023	0\\
-0.986383983587882	-0.19114907059778	0\\
-0.981944911789065	0.219998809200912	0\\
-0.867732709877209	0.43308498159995	0\\
-0.770342430927503	0.589949823503017	0\\
-0.363889282374692	0.918073227675305	0\\
0.361579796769419	0.910807581935862	0\\
0.803796174082175	0.56696048879707	0\\
0.897074638892318	0.375629748091962	0\\
0.986383983587882	0.191149070597781	0\\
}--cycle;

\addplot3[area legend,solid,line width=1.0pt,fill=blue,opacity=5.000000e-01,draw=black,forget plot]
table[row sep=crcr] {%
x	y	z\\
0.663797554314758	-0.444089642980734	0\\
0.652664050689787	-0.467516527092453	0\\
0.643118816247379	-0.484776964005838	0\\
0.604952886488804	-0.553679993377209	0\\
0.564438643665277	-0.590006647804359	0\\
0.536950146861247	-0.608052172014543	0\\
0.510420041002164	-0.625368004104717	0\\
0.442967539417639	-0.636931045535682	0\\
-0.357032460582361	-0.636931045535682	0\\
-0.43057598974838	-0.636930865593846	0\\
-0.470113575200223	-0.628058442294875	0\\
-0.499764109859264	-0.619758689070668	0\\
-0.570080869348126	-0.57546815299064	0\\
-0.615614441543092	-0.538351295163075	0\\
-0.625465582793754	-0.517530645363222	0\\
-0.63490022516507	-0.49745473150767	0\\
-0.663394163708371	-0.400875888430127	0\\
-0.663797554314752	-0.355910357019264	0\\
-0.663797554314752	0.444089642980736	0\\
-0.65266405068971	0.467516527092592	0\\
-0.643118816247379	0.484776964005838	0\\
-0.604952886488804	0.553679993377209	0\\
-0.564438643665277	0.590006647804359	0\\
-0.536950146861256	0.608052172014537	0\\
-0.510420041002164	0.625368004104717	0\\
-0.442967539417639	0.636931045535684	0\\
0.357032460582361	0.636931045535684	0\\
0.43057598974838	0.636930865593848	0\\
0.470113575200118	0.628058442294904	0\\
0.499764109859264	0.619758689070668	0\\
0.570080869348126	0.57546815299064	0\\
0.615614441543092	0.538351295163075	0\\
0.625465582793755	0.517530645363221	0\\
0.63490022516507	0.49745473150767	0\\
0.663394163708371	0.400875888430127	0\\
0.663797554314758	0.355910357019266	0\\
}--cycle;

\end{axis}
\end{tikzpicture}%
		\end{subfigure}
	\end{center}
	\caption{Invariant sets for interconnected rotational systems. The invariant sets $\mathcal X_1$, $\mathcal X_2$ and $\mathcal X_3$ are plotted in red. The sets of possible successor states when using the jointly synthesized feedback controller are depicted in blue. Since the possible successor states are contained inside the red sets, these are indeed robustly controlled invariant. As can be seen, the successor states set for the ``middle'' subsystem is slightly larger because that system is affected by ``disturbance'' from both neighboring subsystems.}
	\label{fig:example_rotation}
\end{figure}

We solved the LMI's in Theorem \ref{thm:main} for a system with three subsystems and parameters $\theta = \pi/4$, $\alpha_i = 0.8$, $\beta_{ij} = 0.1$ for $|i-j| = 1$ and $0$ otherwise, $u_{max, i} = 0.65$, $d_{max, i} = 0.4$ for all $i,j$, with state constraints $\| x_i \|_\infty \leq 1$, and for $Z = [z_1, z_2, \ldots, z_8]^T$, where $z_k$ are randomly chosen unit vectors in $S^1_+$ (c.f. Remark \ref{remark:chooseZ}). The resulting robustly controlled invariant sets that are depicted in Fig. \ref{fig:example_rotation} were computed in 11 seconds. 

Due to the rotational geometry of this problem, the additional set flexibility introduced by the $Z$ matrix is crucial in order to achieve feasibility. Indeed, we were not able to find decoupled invariant sets consisting of linearly transformed hyper boxes using our implementation of the previous work \cite{Tahir:2014cz}.

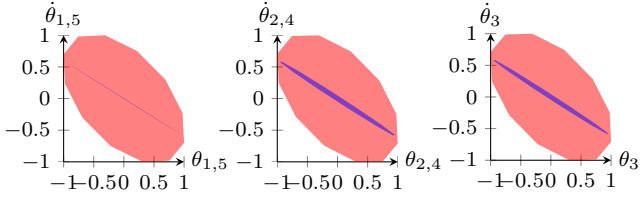
\begin{figure}[tb]
		\vspace{5pt}
		\begin{subfigure}{0.3\columnwidth}
			\setlength\figurewidth{0.65\columnwidth} 
			\setlength\figureheight{0.65\columnwidth} 
			\footnotesize 
%
\begin{tikzpicture}

\begin{axis}[%
width=0.95092\figurewidth,
height=\figureheight,
at={(0\figurewidth,0\figureheight)},
scale only axis,
every outer x axis line/.append style={black},
every x tick label/.append style={font=\color{black}},
xmin=-1,
xmax=1,
xmajorgrids,
every outer y axis line/.append style={black},
every y tick label/.append style={font=\color{black}},
ymin=-1,
ymax=1,
ymajorgrids,
xlabel={$\theta_{1,5}$},
ylabel={$\dot \theta_{1,5}$},
axis x line*=bottom,
axis y line*=left,
xmajorgrids=false, ymajorgrids=false, axis x line=bottom, axis y line=left, every axis x label/.style={at={(current axis.south east)},anchor=west},  every axis y label/.style={at={(current axis.north west)},anchor=south}
]

\addplot[area legend,line width=1.0pt,fill=red,opacity=5.000000e-01,draw=none,forget plot]
table[row sep=crcr] {%
x	y\\
0.975331691723992	-0.229758925821261\\
0.999970303387083	-0.698962509262383\\
0.756667679802498	-0.980879652807021\\
0.310616562476105	-0.999970885509896\\
-0.218664011921491	-0.751120726985761\\
-0.689353740910977	-0.301008376247512\\
-0.975331691723992	0.229758925821261\\
-0.999970303387083	0.698962509262383\\
-0.756667679802498	0.980879652807021\\
-0.310616562476105	0.999970885509896\\
0.218664011921493	0.75112072698576\\
0.689353740910977	0.301008376247512\\
}--cycle;

\addplot[area legend,line width=1.0pt,fill=blue,opacity=5.000000e-01,draw=none,forget plot]
table[row sep=crcr] {%
x	y\\
0.293776084620069	-0.171691811429078\\
0.719454578535728	-0.426697369075795\\
0.952355799141866	-0.567367367182197\\
0.952355799141866	-0.567384863663159\\
0.930074052460845	-0.556026889837347\\
0.658579714521796	-0.3956843039936\\
0.210619473925116	-0.129320772521071\\
-0.293776084620067	0.171691811429077\\
-0.719454578535728	0.426697369075795\\
-0.952355799141866	0.567367367182197\\
-0.952355799141866	0.567384863663159\\
-0.930074052460845	0.556026889837347\\
-0.658579714521796	0.3956843039936\\
-0.210619473925116	0.129320772521071\\
}--cycle;

\end{axis}
\end{tikzpicture}%
		\end{subfigure}~ 
		\begin{subfigure}{0.3\columnwidth}
			\setlength\figurewidth{0.65\columnwidth} 
			\setlength\figureheight{0.65\columnwidth} 
			\footnotesize 
%
\begin{tikzpicture}

\begin{axis}[%
width=0.95092\figurewidth,
height=\figureheight,
at={(0\figurewidth,0\figureheight)},
scale only axis,
every outer x axis line/.append style={black},
every x tick label/.append style={font=\color{black}},
xmin=-1,
xmax=1,
xmajorgrids,
every outer y axis line/.append style={black},
every y tick label/.append style={font=\color{black}},
ymin=-1,
ymax=1,
ymajorgrids,
xlabel={$\theta_{2,4}$},
ylabel={$\dot \theta_{2,4}$},
axis x line*=bottom,
axis y line*=left,
xmajorgrids=false, ymajorgrids=false, axis x line=bottom, axis y line=left, every axis x label/.style={at={(current axis.south east)},anchor=west},  every axis y label/.style={at={(current axis.north west)},anchor=south}
]

\addplot[area legend,line width=1.0pt,fill=red,opacity=5.000000e-01,draw=none,forget plot]
table[row sep=crcr] {%
x	y\\
0.976601436526016	-0.243485550008703\\
0.9999702593325	-0.710850109097116\\
0.7553978586957	-0.987742955513381\\
0.308417211857187	-0.999970874670305\\
-0.221203577830317	-0.744257405504678\\
-0.691553047475312	-0.289120765573191\\
-0.976601436526016	0.243485550008703\\
-0.999970259332499	0.710850109097116\\
-0.755397858695699	0.987742955513381\\
-0.308417211857186	0.999970874670306\\
0.221203577830318	0.744257405504678\\
0.691553047475312	0.289120765573191\\
}--cycle;

\addplot[area legend,line width=1.0pt,fill=blue,opacity=5.000000e-01,draw=none,forget plot]
table[row sep=crcr] {%
x	y\\
0.295629318380786	-0.134095906280387\\
0.720465124032631	-0.408642016000848\\
0.952252881525146	-0.573644558475071\\
0.952252881525146	-0.574004843683985\\
0.928885248422788	-0.585251520906917\\
0.656623563144362	-0.439728794799142\\
0.208420124390156	-0.176429362301612\\
-0.295629318380785	0.134095906280386\\
-0.720465124032631	0.408642016000848\\
-0.952252881525146	0.573644558475071\\
-0.952252881525146	0.574004843683985\\
-0.928885248422788	0.585251520906917\\
-0.656623563144361	0.439728794799141\\
-0.208420124390155	0.176429362301611\\
}--cycle;

\end{axis}
\end{tikzpicture}%
		\end{subfigure}~ 
		\begin{subfigure}{0.3\columnwidth}
			\setlength\figurewidth{0.65\columnwidth} 
			\setlength\figureheight{0.65\columnwidth} 
			\footnotesize 
%
\begin{tikzpicture}

\begin{axis}[%
width=0.95092\figurewidth,
height=\figureheight,
at={(0\figurewidth,0\figureheight)},
scale only axis,
every outer x axis line/.append style={black},
every x tick label/.append style={font=\color{black}},
xmin=-1,
xmax=1,
xmajorgrids,
every outer y axis line/.append style={black},
every y tick label/.append style={font=\color{black}},
ymin=-1,
ymax=1,
ymajorgrids,
xlabel={$\theta_3$},
ylabel={$\dot \theta_3$},
axis x line*=bottom,
axis y line*=left,
xmajorgrids=false, ymajorgrids=false, axis x line=bottom, axis y line=left, every axis x label/.style={at={(current axis.south east)},anchor=west},  every axis y label/.style={at={(current axis.north west)},anchor=south}
]

\addplot[area legend,line width=1.0pt,fill=red,opacity=5.000000e-01,draw=none,forget plot]
table[row sep=crcr] {%
x	y\\
0.976732631079253	-0.243447047018641\\
0.999970253248501	-0.710816759830569\\
0.755266653604668	-0.987723695879389\\
0.308189964057306	-0.999970865272243\\
-0.221465977474585	-0.744276648860749\\
-0.691780289191195	-0.289154105441673\\
-0.976732631079253	0.243447047018641\\
-0.999970253248501	0.710816759830569\\
-0.755266653604667	0.987723695879389\\
-0.308189964057306	0.999970865272243\\
0.221465977474585	0.744276648860749\\
0.691780289191194	0.289154105441674\\
}--cycle;

\addplot[area legend,line width=1.0pt,fill=blue,opacity=5.000000e-01,draw=none,forget plot]
table[row sep=crcr] {%
x	y\\
0.29589364236066	-0.137883223221194\\
0.720695699735361	-0.412064089940538\\
0.952387926377389	-0.575822636820693\\
0.952387926377389	-0.575897872798561\\
0.928888577265444	-0.585355129449296\\
0.656494284016729	-0.437977031588433\\
0.208192877530082	-0.173253421519824\\
-0.29589364236066	0.137883223221193\\
-0.720695699735362	0.412064089940538\\
-0.952387926377389	0.575822636820693\\
-0.952387926377389	0.575897872798561\\
-0.928888577265444	0.585355129449296\\
-0.656494284016728	0.437977031588433\\
-0.208192877530082	0.173253421519824\\
}--cycle;

\end{axis}
\end{tikzpicture}%
		\end{subfigure}
	\caption{Invariant sets (red) for interconnected pendulums at positions 1, 2, and 3, together with possible successor states (blue) when the jointly synthesized feedback controller is used. The sets pertaining to pendulums at positions 4 and 5 are symmetric to those at positions 2 and 1, respectively.}
	\label{fig:example_pendulum}
\end{figure}

Next, we present another example of finding invariant sets, this time for an array of $N$ undisturbed inverted pendulums connected by springs and dampers, an example taken from \cite{conte2012distributed}. A pendulum at position $i$ in the interior of the array (i.e. $i \not \in \{1, N\}$) is described by the states $(\theta_i, \dot \theta_i)$ and has the linearized dynamics
\begin{equation}
	\left[\begin{smallmatrix}
		\dot \theta_i \\
		\ddot \theta_i
	\end{smallmatrix} \right] = \left[\begin{smallmatrix}
		0 & 1 \\ -2 k & -2 c
	\end{smallmatrix}\right] \left[\begin{smallmatrix}
		\theta_i \\
		\dot \theta_i	
	\end{smallmatrix}\right] + \left[\begin{smallmatrix}
	0 \\ u_i
	\end{smallmatrix} \right] +  \left[\begin{smallmatrix}
		0 & 0 \\ k & c	
	\end{smallmatrix} \right] \left[\begin{smallmatrix}
		\theta_{i+1} + \theta_{i-1} \\
		\dot \theta_{i+1} + \dot \theta_{i-1}
	\end{smallmatrix} \right].
\end{equation}
The pendulum at position 1 is only connected to the pendulum at position 2 and has the dynamics
\begin{equation}
	\left[\begin{smallmatrix}
		\dot \theta_1 \\
		\ddot \theta_1
	\end{smallmatrix} \right] = \left[\begin{smallmatrix}
		0 & 1 \\ k & c
	\end{smallmatrix}\right] \left[\begin{smallmatrix}
		\theta_1 \\
		\dot \theta_1	
	\end{smallmatrix}\right]  + \left[\begin{smallmatrix}
	0 \\ u_1
	\end{smallmatrix} \right] + \left[\begin{smallmatrix}
		0 & 0 \\ k & c	
	\end{smallmatrix} \right] \left[\begin{smallmatrix}
		\theta_{2} \\
		\dot \theta_{2}
	\end{smallmatrix} \right],
\end{equation}
and symmetrically for the pendulum at position $N$. We used the same parameter values as in the cited paper which were as follows. The spring and damper parameters are set to $k = c = 3$, and we discretized the continuous time dynamics using a time step $\Delta t = 0.1 \; s$ using Euler forward. We imposed input bounds of $| u_i | < 10$ for all $i$ and state constraints $\max(|\theta_i|, |\dot \theta_i|) \leq 1$ and allowed the input of pendulum $i$ to depend on the states of pendulums $i-1, i$ and $i+1$ (i.e. relaxing $K$ to be a tri-block diagonal matrix). For $N = 5$, and 6 generators per subsystem, we obtained the invariant sets depicted in Fig. \ref{fig:example_pendulum} after finding an initial solution from the LMI's in Theorem \ref{thm:main} and iteratively updating it as described in Section \ref{sub:discussion_about_conservativeness}. The total computation time for initial solution and 5 iterations was 28 s.

\subsection{Local synthesis inside invariant sets} 
\label{sub:robot_uav}

\begin{figure}[tb]
		\begin{subfigure}{0.4\columnwidth}
			\setlength\figurewidth{0.8\columnwidth} 
			\setlength\figureheight{0.8\columnwidth} 
			\footnotesize 
%
\begin{tikzpicture}

\begin{axis}[%
width=0.95092\figurewidth,
height=\figureheight,
at={(0\figurewidth,0\figureheight)},
scale only axis,
every outer x axis line/.append style={black},
every x tick label/.append style={font=\color{black}},
xmin=-0.550317897941747,
xmax=0.550317897941747,
xlabel={$x^1$},
xmajorgrids,
every outer y axis line/.append style={black},
every y tick label/.append style={font=\color{black}},
ymin=-0.388372943289661,
ymax=0.388372943289661,
ylabel={$v_x^1$},
ymajorgrids,
every outer z axis line/.append style={black},
every z tick label/.append style={font=\color{black}},
zmin=-1,
zmax=1,
zmajorgrids,
view={0}{90},
axis x line*=bottom,
axis y line*=left,
axis z line*=left,
xmajorgrids=false, ymajorgrids=false, axis x line=bottom, axis y line=left, every axis x label/.style={at={(current axis.south east)},anchor=west},  every axis y label/.style={at={(current axis.north west)},anchor=south}
]

\addplot3[area legend,solid,line width=1.0pt,fill=red,opacity=5.000000e-01,draw=black,forget plot]
table[row sep=crcr] {%
x	y	z\\
0.429270638715995	-0.0890309540233828	0\\
0.500288998128861	-0.273231136786231	0\\
0.380213964454135	-0.35306631208151	0\\
0.114910119355274	-0.29804215644423	0\\
-0.194285485685995	-0.129176027125568	0\\
-0.429270638715994	0.0890309540233828	0\\
-0.500288998128861	0.273231136786231	0\\
-0.380213964454136	0.35306631208151	0\\
-0.114910119355274	0.29804215644423	0\\
0.194285485685994	0.129176027125568	0\\
}--cycle;

\end{axis}
\end{tikzpicture}%
		\end{subfigure}~
		\begin{subfigure}{0.4\columnwidth}
			\setlength\figurewidth{0.8\columnwidth} 
			\setlength\figureheight{0.8\columnwidth} 
			\footnotesize 
%
\begin{tikzpicture}

\begin{axis}[%
width=0.95092\figurewidth,
height=\figureheight,
at={(0\figurewidth,0\figureheight)},
scale only axis,
every outer x axis line/.append style={black},
every x tick label/.append style={font=\color{black}},
xmin=-0.550315421396435,
xmax=0.550315421396436,
xlabel={$x^2$},
xmajorgrids,
every outer y axis line/.append style={black},
every y tick label/.append style={font=\color{black}},
ymin=-0.388372264909231,
ymax=0.388372264909231,
ylabel={$v_x^2$},
ymajorgrids,
every outer z axis line/.append style={black},
every z tick label/.append style={font=\color{black}},
zmin=-1,
zmax=1,
zmajorgrids,
view={0}{90},
axis x line*=bottom,
axis y line*=left,
axis z line*=left,
xmajorgrids=false, ymajorgrids=false, axis x line=bottom, axis y line=left, every axis x label/.style={at={(current axis.south east)},anchor=west},  every axis y label/.style={at={(current axis.north west)},anchor=south}
]

\addplot3[area legend,solid,line width=1.0pt,fill=red,opacity=5.000000e-01,draw=black,forget plot]
table[row sep=crcr] {%
x	y	z\\
0.429268706050316	-0.0890310122053967	0\\
0.500286746724032	-0.273230791597272	0\\
0.380212254270278	-0.353065695372028	0\\
0.114909603624495	-0.298041503776286	0\\
-0.194284609972067	-0.129175587796133	0\\
-0.429268706050316	0.0890310122053965	0\\
-0.500286746724032	0.273230791597272	0\\
-0.380212254270278	0.353065695372028	0\\
-0.114909603624495	0.298041503776286	0\\
0.194284609972067	0.129175587796133	0\\
}--cycle;

\end{axis}
\end{tikzpicture}%
		\end{subfigure}
	\caption{Invariant sets for toy robot/UAV example.}
	\label{fig:example_uav}
\end{figure}
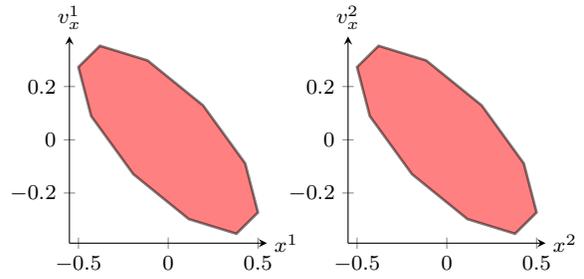

Next, we consider a scenario involving a tethered UAV and a ground vehicle, where each vehicle is given a surveillance task that requires visiting certain regions infinitely often. The tether can be used to power the UAV to significantly increase the duration it is airborne, however it induces dynamic coupling between the UAV and the ground vehicle. This coupling is modeled as a spring. For simplicity, the vehicles are modeled as double integrators and their motion is constrained to one dimension:
\begin{equation}
	\left[ \begin{smallmatrix}
		x^1(t+1) \\ v_x^1(t+1) \\ x^2(t+1) \\ v_x^2(t+1)
	\end{smallmatrix}\right] = 
	\left[\begin{smallmatrix}
		1 & 1 & 0 & 0 \\ k & 1 & -k & 0 \\ 0 & 0 & 1 & 1 \\ -k & 0 & k & 1
	\end{smallmatrix} \right]
\left[ \begin{smallmatrix}
		x^1(t) \\ v_x^1(t) \\ x^2(t) \\ v_x^2(t)
	\end{smallmatrix}\right]
	+ \left[ \begin{smallmatrix}
		0 & 0 \\ 1 & 0 \\ 0 & 0 \\ 1 & 0
	\end{smallmatrix}\right] \left[\begin{smallmatrix}
		u^1 \\ u^2
	\end{smallmatrix}\right].
\end{equation}
Our objective is to find separable invariant sets for the two subsystems $(x^1, v_x^1)$ and $(x^2, v_x^2)$ and then let each system perform additional control objectives while still guaranteeing overall invariance. Using a small coupling $k = 0.1$ and a bound $\| u \|_\infty \leq 0.3$, we could in 1.7 s compute the invariant sets depicted in Fig. \ref{fig:example_uav} that consist of 5 zonotope generators each.

We now illustrate how additional control objectives can be modularly realized inside these sets. Let $\mathcal X_1, \mathcal X_2$ be the invariant set pertaining to the subsystem $(x^1, v_x^1)$. We consider a surveillance-like task where the ground robot is required to visit the goal sets $G_{1}^+, G_{1}^- \subset \mathcal X_1$ infinitely often; whereas the UAV is required to visit $G_{2}^+, G_{2}^- \subset \mathcal X_2$ infinitely often, where the goal sets are taken to be $G_1^\pm = \{ (\pm x^1, v_x^1) : x^1 \in [0.2, 0.35] \}$ and $G_{2}^{\pm} = \{ (\pm x^2, v_x^2) : x^2 \in [0.05, 0.18] \}$. The local objectives can be expressed in LTL as $\varphi_i = \square \lozenge G_i^+ \land \square \lozenge G_i^-$ for $i=1,2$.

\begin{figure}[tb]
		\vspace{5pt}
		\begin{subfigure}{0.4\columnwidth}
			\setlength\figurewidth{0.9\columnwidth} 
			\setlength\figureheight{0.9\columnwidth} 
			\footnotesize 
%
\definecolor{mycolor1}{rgb}{0.00000,0.50196,0.00000}%
\begin{tikzpicture}

\begin{axis}[%
width=0.95092\figurewidth,
height=\figureheight,
at={(0\figurewidth,0\figureheight)},
scale only axis,
every outer x axis line/.append style={black},
every x tick label/.append style={font=\color{black}},
xmin=-0.550138295593453,
xmax=0.550138295593453,
xlabel={$x^1$},
xmajorgrids,
every outer y axis line/.append style={black},
every y tick label/.append style={font=\color{black}},
ymin=-0.388306702613982,
ymax=0.388306702613982,
ylabel={$v_x^1$},
ymajorgrids,
axis x line*=bottom,
axis y line*=left,
xmajorgrids=false, ymajorgrids=false, axis x line=bottom, axis y line=left, every axis x label/.style={at={(current axis.south east)},anchor=west},  every axis y label/.style={at={(current axis.north west)},anchor=south}
]

\addplot[area legend,solid,line width=1.0pt,fill=white,draw=black,forget plot]
table[row sep=crcr] {%
x	y\\
0.429114561074106	-0.0890241471376553\\
0.500125723266776	-0.273189712636759\\
0.380105857819661	-0.353006093285438\\
0.114898474008371	-0.297986144534896\\
-0.194196221618621	-0.129145616748562\\
-0.429114561074106	0.0890241471376554\\
-0.500125723266776	0.273189712636759\\
-0.380105857819661	0.353006093285438\\
-0.11489847400837	0.297986144534895\\
0.194196221618621	0.129145616748562\\
}--cycle;

\addplot[area legend,line width=1.0pt,fill=mycolor1,opacity=1.000000e-01,draw=none,forget plot]
table[row sep=crcr] {%
x	y\\
-0.2	-0.12375562053946\\
-0.35	0.0155500831309578\\
-0.35	0.346760329194995\\
-0.2	0.315641315108819\\
}--cycle;

\addplot[area legend,line width=1.0pt,fill=mycolor1,opacity=1.000000e-01,draw=none,forget plot]
table[row sep=crcr] {%
x	y\\
0.0776267735560436	-0.277626773556044\\
-0.194196221618621	-0.129145616748561\\
-0.429114561074106	0.0890241471376554\\
-0.500125723266776	0.273189712636759\\
-0.483948195742831	0.28394819574283\\
}--cycle;

\addplot[area legend,line width=1.0pt,fill=mycolor1,opacity=1.000000e-01,draw=none,forget plot]
table[row sep=crcr] {%
x	y\\
0.165755711392173	-0.308536991778432\\
0.114898474008371	-0.297986144534895\\
-0.194196221618621	-0.129145616748561\\
-0.429114561074106	0.0890241471376553\\
-0.500125723266776	0.273189712636759\\
-0.478005661068942	0.287900138263196\\
-0.372840677914528	0.230059397528268\\
}--cycle;

\addplot[area legend,line width=1.0pt,fill=mycolor1,opacity=1.000000e-01,draw=none,forget plot]
table[row sep=crcr] {%
x	y\\
0.324863988369438	-0.34154560986165\\
0.114898474008371	-0.297986144534896\\
-0.194196221618621	-0.129145616748562\\
-0.429114561074106	0.0890241471376553\\
-0.500125723266776	0.273189712636759\\
-0.454459374755658	0.30355904059776\\
-0.156195568831914	0.139513947339701\\
}--cycle;

\addplot[area legend,line width=1.0pt,fill=mycolor1,opacity=1.000000e-01,draw=none,forget plot]
table[row sep=crcr] {%
x	y\\
0.483150827009774	-0.284478467011488\\
0.380105857819658	-0.353006093285438\\
0.114898474008371	-0.297986144534896\\
-0.194196221618621	-0.129145616748561\\
-0.429114561074106	0.0890241471376553\\
-0.500125723266776	0.273189712636759\\
-0.402567639107423	0.338068415438254\\
0.182258102375814	0.0164142576224736\\
}--cycle;

\addplot[area legend,line width=1.0pt,fill=mycolor1,opacity=1.000000e-01,draw=none,forget plot]
table[row sep=crcr] {%
x	y\\
0.429114561074106	-0.0890241471376555\\
0.500125723266775	-0.27318971263676\\
0.38010585781966	-0.353006093285438\\
0.114898474008371	-0.297986144534895\\
-0.194196221618621	-0.129145616748562\\
-0.429114561074106	0.0890241471376554\\
-0.500125723266776	0.273189712636759\\
-0.380105857819661	0.353006093285438\\
-0.177450079791988	0.310963106544851\\
-0.0935002273302435	0.275758329706055\\
0.224880642055877	0.100648851543689\\
}--cycle;

\addplot[area legend,line width=1.0pt,fill=mycolor1,opacity=1.000000e-01,draw=none,forget plot]
table[row sep=crcr] {%
x	y\\
0.429114561074106	-0.0890241471376553\\
0.500125723266776	-0.273189712636759\\
0.380105857819661	-0.353006093285438\\
0.114898474008371	-0.297986144534896\\
-0.194196221618621	-0.129145616748562\\
-0.429114561074106	0.0890241471376554\\
-0.500125723266776	0.273189712636759\\
-0.380105857819661	0.353006093285438\\
-0.11489847400837	0.297986144534895\\
0.194196221618621	0.129145616748562\\
}--cycle;

\addplot[area legend,solid,line width=1.0pt,fill=mycolor1,draw=black,forget plot]
table[row sep=crcr] {%
x	y\\
-0.2	-0.12375562053946\\
-0.35	0.0155500831309578\\
-0.35	0.346760329194995\\
-0.2	0.315641315108819\\
}--cycle;

\addplot[area legend,solid,line width=1.0pt,fill=blue,draw=black,forget plot]
table[row sep=crcr] {%
x	y\\
0.35	-0.346760329194995\\
0.2	-0.315641315108819\\
0.2	0.12375562053946\\
0.35	-0.0155500831309578\\
}--cycle;

\end{axis}
\end{tikzpicture}%
		\end{subfigure}~
		\begin{subfigure}{0.4\columnwidth}
			\setlength\figurewidth{0.9\columnwidth} 
			\setlength\figureheight{0.9\columnwidth} 
			\footnotesize 
%
\definecolor{mycolor1}{rgb}{0.00000,0.50196,0.00000}%
\begin{tikzpicture}

\begin{axis}[%
width=0.95092\figurewidth,
height=\figureheight,
at={(0\figurewidth,0\figureheight)},
scale only axis,
every outer x axis line/.append style={black},
every x tick label/.append style={font=\color{black}},
xmin=-0.550138295593453,
xmax=0.550138295593453,
xlabel={$x^2$},
xmajorgrids,
every outer y axis line/.append style={black},
every y tick label/.append style={font=\color{black}},
ymin=-0.388306702613982,
ymax=0.388306702613982,
ylabel={$v_x^2$},
ymajorgrids,
axis x line*=bottom,
axis y line*=left,
xmajorgrids=false, ymajorgrids=false, axis x line=bottom, axis y line=left, every axis x label/.style={at={(current axis.south east)},anchor=west},  every axis y label/.style={at={(current axis.north west)},anchor=south}
]

\addplot[area legend,solid,line width=1.0pt,fill=white,draw=black,forget plot]
table[row sep=crcr] {%
x	y\\
0.429114561074106	-0.0890241471376553\\
0.500125723266776	-0.273189712636759\\
0.380105857819661	-0.353006093285438\\
0.114898474008371	-0.297986144534896\\
-0.194196221618621	-0.129145616748562\\
-0.429114561074106	0.0890241471376554\\
-0.500125723266776	0.273189712636759\\
-0.380105857819661	0.353006093285438\\
-0.11489847400837	0.297986144534895\\
0.194196221618621	0.129145616748562\\
}--cycle;

\addplot[area legend,line width=1.0pt,fill=mycolor1,opacity=1.000000e-01,draw=none,forget plot]
table[row sep=crcr] {%
x	y\\
-0.05	-0.207896783036872\\
-0.18	-0.136880381322429\\
-0.18	0.311474076003298\\
-0.11486688245708	0.297960265759793\\
-0.05	0.262524784355673\\
}--cycle;

\addplot[area legend,line width=1.0pt,fill=mycolor1,opacity=1.000000e-01,draw=none,forget plot]
table[row sep=crcr] {%
x	y\\
0.282803792740626	-0.332803792740626\\
0.118770106053092	-0.298770106053092\\
-0.471898809645601	0.2918988096456\\
-0.393819786441694	0.343819786441694\\
}--cycle;

\addplot[area legend,line width=1.0pt,fill=mycolor1,opacity=1.000000e-01,draw=none,forget plot]
table[row sep=crcr] {%
x	y\\
0.457839964564442	-0.301247658277229\\
0.380045408951279	-0.352979469858747\\
0.172643825707491	-0.309947817975781\\
-0.249467823400728	-0.0777864109662603\\
-0.429067868954228	0.0890229422660751\\
-0.500059506494442	0.273172513802578\\
-0.416227721980958	0.328918960926189\\
0.125774649890114	0.030817656397099\\
}--cycle;

\addplot[area legend,line width=1.0pt,fill=mycolor1,opacity=1.000000e-01,draw=none,forget plot]
table[row sep=crcr] {%
x	y\\
0.429067868954228	-0.0890229422660754\\
0.500059506494443	-0.273172513802578\\
0.380045408951279	-0.352979469858747\\
0.11486688245708	-0.297960265759793\\
-0.194186888953985	-0.129130367437549\\
-0.429067868954228	0.0890229422660751\\
-0.500059506494442	0.273172513802578\\
-0.380045408951279	0.352979469858747\\
-0.228122342485892	0.321458492984396\\
-0.172256037978115	0.298030687868231\\
0.280366429705677	0.0490883306421456\\
}--cycle;

\addplot[area legend,line width=1.0pt,fill=mycolor1,opacity=1.000000e-01,draw=none,forget plot]
table[row sep=crcr] {%
x	y\\
0.429067868954228	-0.0890229422660751\\
0.500059506494442	-0.273172513802578\\
0.380045408951278	-0.352979469858747\\
0.11486688245708	-0.297960265759793\\
-0.194186888953984	-0.129130367437549\\
-0.429067868954228	0.0890229422660751\\
-0.500059506494442	0.273172513802578\\
-0.380045408951279	0.352979469858746\\
-0.114866882457079	0.297960265759793\\
0.194186888953984	0.12913036743755\\
}--cycle;

\addplot[area legend,line width=1.0pt,fill=mycolor1,opacity=1.000000e-01,draw=none,forget plot]
table[row sep=crcr] {%
x	y\\
0.429067868954228	-0.0890229422660751\\
0.500059506494442	-0.273172513802578\\
0.380045408951278	-0.352979469858747\\
0.11486688245708	-0.297960265759793\\
-0.194186888953984	-0.129130367437549\\
-0.429067868954228	0.0890229422660751\\
-0.500059506494442	0.273172513802578\\
-0.380045408951279	0.352979469858746\\
-0.114866882457079	0.297960265759793\\
0.194186888953984	0.12913036743755\\
}--cycle;

\addplot[area legend,solid,line width=1.0pt,fill=mycolor1,draw=black,forget plot]
table[row sep=crcr] {%
x	y\\
-0.05	-0.207896783036872\\
-0.18	-0.136880381322429\\
-0.18	0.311474076003298\\
-0.11486688245708	0.297960265759793\\
-0.05	0.262524784355673\\
}--cycle;

\addplot[area legend,solid,line width=1.0pt,fill=blue,draw=black,forget plot]
table[row sep=crcr] {%
x	y\\
0.18	-0.311474076003298\\
0.11486688245708	-0.297960265759793\\
0.05	-0.262524784355673\\
0.05	0.207896783036871\\
0.18	0.136880381322429\\
}--cycle;

\end{axis}
\end{tikzpicture}%
		\end{subfigure}
	\caption{Illustration of a simple synthesis problem inside one of the invariant sets. For $i=1$ (left) and $i=2$ (right), the goal sets $G_i^-, G_i^+$ are shown in green, blue. The light green sets depict regions from where $G_i^-$ is reachable.}
	\label{fig:reach}
\end{figure}
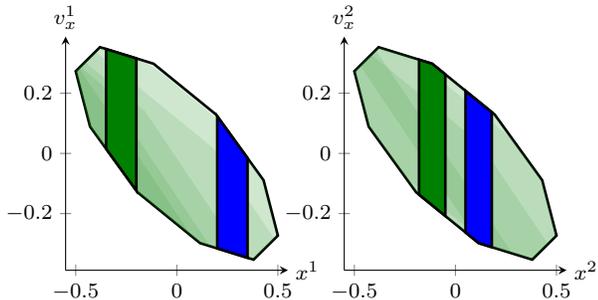

\begin{figure}[tb]
	\setlength\figurewidth{0.8\columnwidth} 
	\setlength\figureheight{0.4\columnwidth} 
	\footnotesize \input{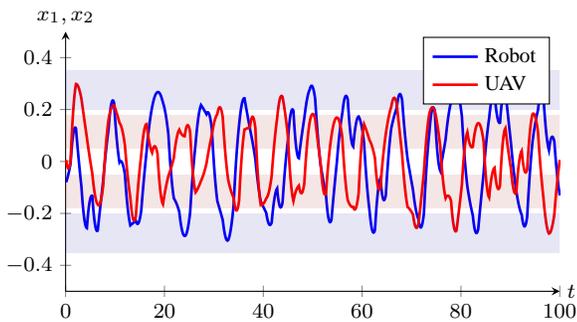}
	\caption{Subsystem trajectories that visit given regions (marked with red, blue) infinitely often.}
	\label{fig:reach_trajs}
\end{figure}

We use reachability computations similar to \cite{nilsson2014preliminary} to synthesize the local controllers, but any LTL synthesis method can be used in this step. Fig. \ref{fig:reach} depicts $G_i^\pm$, together with robust (with respect to ``disturbance'' induced from the other subsystem) backwards reachable sets of $G_i^-$ contained inside $\mathcal X_i$ in lighter green (the backwards reachable sets from $G_i^+$ are symmetric). Since the sets from where $G_i^-$ is reachable eventually cover $G_i^+$, together with the fact that $G_i^+$ is reachable from $G_i^-$ by symmetry, the specification is realizable. Fig. \ref{fig:reach_trajs} shows trajectories of a simulation where both systems satisfy their control objectives, while simultaneously countering the ``disturbance'' they cause each other. 

\begin{figure}[tb]
	\setlength\figurewidth{0.8\columnwidth} 
	\setlength\figureheight{0.4\columnwidth} 
	\footnotesize \input{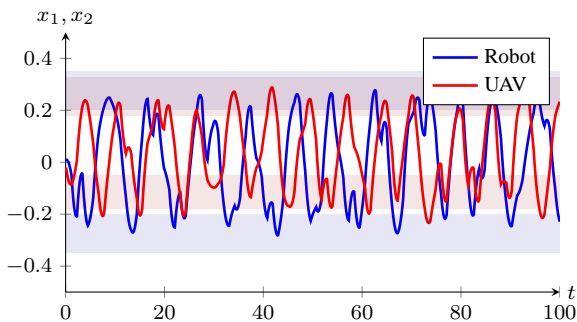}
	\caption{Re-synthesis can be done separately for the different subsystems, thanks to the inherent robust modularity. The plot shows the same simulation as in Fig. \ref{fig:reach_trajs}, but with a different local control objective for the UAV.}
	\label{fig:reach_trajs_resynth}
\end{figure}

Now, assume the task specification for the UAV changes to $\varphi'_2= \square \lozenge G'_2 \land \square \lozenge G''_2$, with $G'_2 = \{ (x^2, v_x^2) : x^2\in[-0.18, -0.05]\}$ and $G''_2 = \{ (x^2, v_x^2) : x^2 \in [0.18, 0.33]\}$. By construction of the invariant sets, if a new controller can be synthesized for the UAV, the new specification $\varphi_1\wedge \varphi'_2$ is guaranteed to be satisfied without making any changes to the controller of the ground robot, despite the potentially different ``disturbance'' inputs it gets from the UAV. Fig. \ref{fig:reach_trajs_resynth} shows the trajectories of a simulation of this new scenario where both systems satisfy their control objectives.

An interesting feature of this problem is that if the invariant sets are increased in size, these control objectives can no longer be realized using this kind of decentralized controllers. As the sizes of the sets increase, more control effort must be reserved for countering the increasing ``disturbance'' from the other subsystem and hence there is less freedom to pursue more sophisticated control objectives.

\section{Conclusions}
\label{sec:conclusions}

In this paper, we proposed a two-step approach for modular control design for dynamically coupled linear subsystems with local state and input constraints. The idea is to first compute a separable controlled invariant set consisting of local robust controlled invariant sets for each subsystem to ``decouple" the subsystem dynamics, and then to use these local invariant sets to separately synthesize local controllers to achieve local tasks. By construction, this approach guarantees correctness when the local controllers are composed together. We proposed LMI based sufficient conditions for computation of the separable controlled invariant set. The proposed invariant set computation scheme is quite flexible as it allows for (i) incorporating external disturbances, (ii) handling different communication constraints (i.e., which states each local controller has access to), and (iii) tuning the complexity of the invariant set description. As demonstrated by examples, this flexibility allows us to find invariant sets even when known polytopic invariant set construction schemes fail.  

There are several directions for future work. On the theoretical side, we are interested in characterizing systems that admit separable invariant sets. There are recent results on characterizations of systems with separable Lyapunov functions \cite{ito2014max}. We are working on extending some of these results to the controlled setting considered in this paper. Automatically decomposing a complex system into subsystems that admit local invariant sets is another direction. On the computational side, we used off-the-shelf convex optimization solvers to compute the invariant sets. It is possible to exploit the structure in the LMI formulation to increase the efficiency of the computation. In particular, we are currently working on implementing our solution using alternating direction method of multipliers \cite{boyd2011distributed}, which will allow the invariant set computation to be performed in a distributed manner as well.

\section*{Acknowlegments}

The authors would like to thank Jessy Grizzle for helpful discussions. The work of PN was supported by NSF grant CNS-1239037. The work of NO was partly supported by NSF grant  CNS-1446298.

\appendix

\subsection{Proof of Lemma \ref{corr:result1b}} 
We prove that the following two statements are equivalent
\begin{enumerate}
	\item 
	\begin{equation}
		\label{eq:lemma1cond1}
		\left[ \begin{smallmatrix}
			X & Y \\
			* & Z
		\end{smallmatrix} \right] \succ 0
	\end{equation}		
	\item $X \succ 0$ and there exists $\Psi$ (full) such that
	\begin{equation}
		\label{eq:lemma1cond2}
		\left[ \begin{smallmatrix}
			\Psi^T + \Psi - X^{-1} & \Psi^T Y \\ * & Z
		\end{smallmatrix} \right] \succ 0.
	\end{equation}
\end{enumerate}

Then Lemma \ref{corr:result1b} follows by replacing $X \mapsto C^T X C$ in \eqref{eq:lemma1cond2}, the congruency transform $\blkdiag{ \{C, I \}}$ and re-defining $C \Psi^T \mapsto \Psi^T$.

($1 \rightarrow 2$): First note that \eqref{eq:lemma1cond1} implies that $X \succ 0$ by a Schur complement argument. By a congruency transform, \eqref{eq:lemma1cond1} is equivalent to
	\begin{equation}
	\label{eq:lemma1cond1_aftercong}
		\forall \; \Psi \; \text{non-singular}, \quad \left[ \begin{smallmatrix}
			\Psi^T X \Psi & \Psi^T Y \\ * & Z
		\end{smallmatrix} \right] \succ 0.
	\end{equation}
	Since $X$ is invertible we can write $\Psi^T X \Psi = \Psi^T + \Psi - X^{-1} + (\Psi-X)^T X^{-1} (\Psi-X)$. Substituting this expression in \eqref{eq:lemma1cond1_aftercong} and choosing $\Psi$ as the (non-singular) matrix $X$ which eliminates the last term shows that there exists $\Psi$ such that \eqref{eq:lemma1cond2} holds.

	($2 \rightarrow 1$): Assuming that \eqref{eq:lemma1cond2} holds, we can ``add back'' the positive term $(\Psi-X)^T X^{-1} (\Psi-X)$ to the top left entry to obtain that
	\begin{equation}
		\exists \Psi \; \text{s.t.} \; \left[ \begin{smallmatrix}
			\Psi ^T X \Psi &  \Psi^T Y \\ * & Z
		\end{smallmatrix} \right] \succ 0.
	\end{equation}
	A contradiction argument shows that $\Psi$ is non-singular, so we can apply a congruency transform that eliminates $\Psi$.

\subsection{Proof of Lemma \ref{lemma:result2}} 
\label{sub:proof_of_lemma_ref_lemma_result2}

	Adding the positive definite terms $X \Gamma X^T$ to the top left block and $(\Theta-\Delta)^T \Delta^{-1}(\Theta - \Delta)$ to the middle block of \eqref{eq:lemma2cond1} preserves positive definiteness. Using the identity $\Theta^T \Delta^{-1} \Theta = \Theta + \Theta^T - \Delta + (\Theta - \Delta)^T \Delta^{-1} (\Theta - \Delta)$ then implies that
	\begin{equation}
		\left[\begin{smallmatrix}
			Z + \Xi + X \Gamma X^T & \left[\begin{smallmatrix}
				-X & I
			\end{smallmatrix}\right] \Theta & V \\
			* & \Theta^T \Delta^{-1} \Theta & 0 \\ * & * & W
		\end{smallmatrix}\right] \succ 0.
	\end{equation}
	A contradiction argument shows that $\Theta$ is non-singular. The congruency transform $\left[\begin{smallmatrix}
	I & 0 & 0 \\ 0 & 0 & I \\ 0 & \Theta^{-T} & 0
	\end{smallmatrix} \right]$ then gives
	\begin{equation}
		\left[\begin{smallmatrix}
			Z + \Xi + X \Gamma X^T&  V & \left[\begin{smallmatrix}
				-X & I
			\end{smallmatrix}\right] \\
			* & W & 0 \\ * & * & \Delta^{-1}
		\end{smallmatrix}\right] \succ 0.
	\end{equation}
	Applying a Schur complement finally implies that
	\begin{equation}
		\begin{aligned}
			0 \prec \left[\begin{smallmatrix}
				Z + \Xi + X \Gamma X^T  & V \\ * & W
			\end{smallmatrix}\right] - \left[\begin{smallmatrix}
			-X & I
			\end{smallmatrix} \right] \Delta \left[\begin{smallmatrix}
			X^T \\ I
			\end{smallmatrix} \right] \\
			= \left[\begin{smallmatrix}
			Z + X Y + Y^T X^T & V \\ * & W
			\end{smallmatrix} \right].
		\end{aligned}
	\end{equation}

\subsection{Proof of Theorem \ref{thm:main}} 
We verify that the satisfaction of the LMI's given in the theorem statement guarantee \eqref{eq:invariance} - \eqref{eq:contconstr}, starting with \eqref{eq:invariance}. Because of symmetry, $x \in \mathcal X$ if and only if $-x \in \mathcal X$, and similarly for $\mathcal D$. Therefore it follows that \eqref{eq:invariance} holds if and only if
\begin{equation}
\label{eq:invoneside}
	e_j^T Z H_x( A_K  x + E d) - 1 \leq 0
\end{equation}
for all $x \in \mathcal X$, $d \in \mathcal D$ and for all $j = 1, \ldots \mathcal N_x$. Furthermore note that $x \in \mathcal X$ if and only if for all diagonal $D_x \succ 0$
\begin{equation}
\label{eq:quadratic}
	(\mathds{1} - Z H_x x)^T D_x (\mathds{1} + Z H_x x) \geq 0.
\end{equation}
Similarly, $d \in \mathcal D$ if and only if for all diagonal $D_d \succ 0$
\begin{equation}
\label{eq:quadratic_d}
	(\mathds{1} - H_d d)^T D_d (\mathds{1} + H_d d) \geq 0.
\end{equation}
The next step is to employ the S Procedure. To prepare for this, we express the left hand side of \eqref{eq:invoneside} in terms of the quadratic forms in \eqref{eq:quadratic}-\eqref{eq:quadratic_d} and an additional quadratic term:
\begin{equation}
\begin{aligned}
	e_j^T & Z H_x A_K x + e_j^T Z H_x E d - 1 \\
	&= -(\mathds{1} - Z H_x x)^T \tilde D_x^j (\mathds{1} + Z H_x x) \\
	& -(\mathds{1} - Z H_d d)^T \tilde D_d^j (\mathds{1} + Z H_d d) \\
	& - \begin{bmatrix}
		x^T & d^T & 1
	\end{bmatrix} L_x^j(\tilde D_x^j, \tilde D_d^j) \begin{bmatrix}
		x^T & d^T & 1
	\end{bmatrix}^T,
\end{aligned}
\end{equation}
for $\tilde D_x^j \succ 0$ (diag), $\tilde D_d^j \succ 0$ (diag) and
\begin{equation}
	\hspace{-1.5mm} {\tiny L^j_x(\tilde D_x^j, \tilde D_d^j) = \left[\begin{smallmatrix}
		H_x^T Z^T \tilde D_x^j Z H_x & 0 & - \frac{1}{2}A_K ^T H_x^T Z^T e_j \\
		* & H_d^T \tilde D_d^j  H_d & -\frac{1}{2} E^T H_x^T Z^T e_j \\
		* & * & 1 - \mathds{1}^T \tilde D_x^j \mathds{1}- \mathds{1}^T \tilde D_d^j \mathds{1}
	\end{smallmatrix}\right].}
\end{equation}
An application of the S procedure \cite{Polik:2007kv} shows that \eqref{eq:invariance} holds if and only if for all $j = 1, \ldots \mathcal N_x$ there exist $\tilde D_x^j \succ 0$ (diag) and $\tilde D_d^j \succ 0$ (diag) such that $L^j_x(\tilde D_x^j, \tilde D_d^j) \succ 0$.

Next, we use Lemma \ref{lemma:thlemma} which implies that \eqref{eq:invariance} holds if and only if for all $j = 1, \ldots \mathcal N_x$ there exist $\tilde D_x^j \succ 0$ (diag), $\tilde D_d^j \succ 0$ (diag), and $\Phi_j$ (sym) such that
\begin{align}
	\label{eq:m1} 
	 M_1 & \dot = \left[\begin{smallmatrix}
		\begin{smallmatrix}
			H_x^T Z^T \tilde D_x^j Z H_x & 0 \\ *  & H_d^T \tilde D_d^j H_d
		\end{smallmatrix} & \begin{smallmatrix}
			-\frac{1}{2} A_K^T & 0 \\ 
			0 & -\frac{1}{2} E^T 
		\end{smallmatrix} \\
		*  & \left[ \tilde \Phi_j^{-1} \right]
	\end{smallmatrix}\right] \succ 0, \\
	\label{eq:m2} 
	M_2 & \dot = \left[\begin{smallmatrix}
		\tilde \Phi_j & \left[\begin{smallmatrix}
			H_x^T Z^T e_j\\
			H_x^T Z^T e_j\\
		\end{smallmatrix}\right] \\
		* & 1-\mathds{1}^T \tilde D_x^j \mathds{1} - \mathds{1}^T \tilde D_d^j \mathds{1}
	\end{smallmatrix}\right] \succ 0.
\end{align}
In the remaining part of the proof, these two matrix inequalities are turned into LMI's.

\paragraph{Treatment of $M_2$}
Let $\bar H_x = \blkdiag{ \{H_x, H_x \} }$ and apply the congruency transform $\blkdiag{ \{ \bar H_x^{-T}, I \} }$ on $M_2$ to obtain the equivalent condition
\begin{equation}
\label{eq:m2eq1}
	 \left[\begin{smallmatrix}
		\bar H_x^{-T} \tilde \Phi_j \bar H_x^{-1} & \left[\begin{smallmatrix}
			Z^T e_j\\
			Z^T e_j\\
		\end{smallmatrix}\right] \\ * & 1-\mathds{1}^T \tilde D_x^j \mathds{1} - \mathds{1}^T \tilde D_d^j \mathds{1}
	\end{smallmatrix}\right] \succ 0.
\end{equation}
Multiply the matrix in \eqref{eq:m2eq1} with a scalar $\lambda_{i(j)} > 0$, where $i(j)$ is the index of the subsystem corresponding to the $j$th inequality in $Z$, and redefine $\Phi_j = \lambda_{i(j)}  \tilde \Phi_j$, $D_x^j = \lambda_{i(j)} \tilde D_x^j$, $D_d^j = \lambda_{i(j)} \tilde  D_d^j$. Note that since $Z^T$ is block diagonal, we have for $\Lambda = \blkdiag{ \{ I_{n_i} \lambda_{i} \}_{i=1}^d }$ that $\Lambda Z^T e_j = \lambda_{i(j)} Z^T e_j$. This results in
\begin{equation}
	 \left[\begin{smallmatrix}
		\bar H_x^{-T}  \Phi_j \bar H_x^{-1} & \left[\begin{smallmatrix} \Lambda Z^T e_j \\ \Lambda Z^T e_j \end{smallmatrix}\right] \\ 
		* & \lambda_j-\mathds{1}^T  D_x^j \mathds{1} -\mathds{1}^T D_d^j \mathds{1}
	\end{smallmatrix}\right] \succ 0.
\end{equation}
For $\bar \Lambda = \blkdiag{ \{ \Lambda, \Lambda \} }$ apply the congruency transform $\blkdiag{ \{\bar \Lambda^{-T}, I \}}$:
\begin{equation}
	 \left[\begin{smallmatrix}
		\bar \Lambda^{-T} \bar H_x^{-T}  \Phi_j \bar H_x ^{-1} \Lambda^{-1} & \left[\begin{smallmatrix}  Z^T e_j \\  Z^T e_j \end{smallmatrix}\right] \\ 
		* & \lambda_j-\mathds{1}^T  D_x^j \mathds{1} - \mathds{1}^T D_d^j \mathds{1}
	\end{smallmatrix}\right] \succ 0.
\end{equation}
We now apply Lemma \ref{corr:result1b} which implies that \eqref{eq:m2} holds if and only if there exist $ \Phi_j$ (sym) and $\Psi_j$ (full) such that
\begin{equation}
\label{eq:pre_result1b}
	{\small \left[\begin{smallmatrix}
			\Psi_j^T \bar \Lambda^{-T} \bar H_x^{-T} + \bar H_x^{-1} \bar \Lambda^{-1} \Psi_j  -  \Phi_j^{-1} & \Psi_j^T \left[\begin{smallmatrix}
				Z^T e_j  \\ Z^T e_j 
			\end{smallmatrix}\right]\\ 
			* & 1-\mathds{1}^T D_x^j \mathds{1} - \mathds{1}^T D_d^j \mathds{1}
		\end{smallmatrix}\right] \succ 0.}
\end{equation}
Finally, we use Lemma \ref{lemma:result2} to obtain the \emph{necessary} condition
\begin{equation}
\label{eq:post_result1b}
\begin{aligned}
{	 \small \Delta_j \dot = \left[\begin{smallmatrix}
		\Gamma_j & \Psi_j \\ * & \Xi_j
	\end{smallmatrix}\right]  \succ 0,}\\
	{ \small \left[\begin{smallmatrix}
				- \tilde \Phi_j^{-1} + \Xi_j & \left[\begin{smallmatrix}
					-\bar H^{-1} \bar \Lambda^{-1} & I
				\end{smallmatrix}\right] \Theta_j &  \Psi_j^T \left[\begin{smallmatrix}
				Z^T e_j  \\ Z^T e_j 
			\end{smallmatrix}\right] \\
				* & \Theta_j + \Theta_j^T - \Delta_j & 0 \\ 
				* & * & 1-\mathds{1}^T D_x^j \mathds{1} - \mathds{1}^T D_d^j \mathds{1}
			\end{smallmatrix}\right]  \succ 0.}
\end{aligned}
\end{equation}
By further restricting to $ \Theta_j = \left[\begin{smallmatrix}
\bar \Lambda & \bar \Lambda \\ \Omega^1_j & \Omega^2_j
\end{smallmatrix}\right]$, we obtain the LMIs in \eqref{eq:lmi_first}.

\paragraph{Treatment of $M_1$}
Apply the congruency transform $\blkdiag{ \{ \lambda_{i(j)} H_x^{-T}, \lambda_{i(j)} H_d^{-T}, I \lambda_{i(j)}^{-1} \} }$ to obtain
\begin{equation}
{ \small \left[\begin{smallmatrix}
		\left[\begin{smallmatrix}
			Z^T \lambda_{i(j)} \tilde D_x^j Z & 0 \\ *  & \lambda_{i(j)} \tilde D_d^j
		\end{smallmatrix}\right] & \left[\begin{smallmatrix}
			-\frac{1}{2} H_x^{-T} A_K^T & 0 \\ 
			0 & -\frac{1}{2} H_d^{-T} E^T 
		\end{smallmatrix}\right] \\
		* & \lambda_{i(j)}^{-1}  \tilde \Phi_j^{-1}
	\end{smallmatrix}\right] \succ 0.}
\end{equation}
Note that $H_x^{-T} A_K^T = ( (A+BK)H_x^{-1} )^T =( AH_x^{-1} + B \hat K)^T$, for $\hat K = K H_x^{-1}$. Thus after the same re-definitions as above for $\tilde D_x^j$, $\tilde D_d^j$ and $\tilde \Phi_j$, we get the LMI \eqref{eq:lmi_third}.

We move on to finding an expression that ensures \eqref{eq:stateconst}. As before, we write for $k = 1, \ldots \mathcal N_s$
\begin{equation}
\begin{aligned}
	e_k^T & \left( H_s x - h_s \right) \\
	&= -(\mathds{1} - Z H_x x)^T D_s^k (\mathds{1} + Z H_x x) \\
	& - \begin{bmatrix}
		x^T & 1
	\end{bmatrix} L_s^k(D_s^k) \begin{bmatrix}
		x^T & 1
	\end{bmatrix}^T,
\end{aligned}
\end{equation}
for 
\begin{equation}
	L_s^k(D_s^k) = \left[\begin{smallmatrix}
		H_x^T Z^T D_s^k Z H_x & - \frac{1}{2} H_s^T e_k \\ * & e_k^T h_s - \mathds{1}^T D_s^k \mathds{1}
	\end{smallmatrix}\right].
\end{equation}
The S procedure applies in the same way as before and by the congruency transform $\blkdiag{ \{ H_x^{-T}, I\} }$ we get the LMI \eqref{eq:lmi_fourth}
which represents necessary and sufficient conditions for \eqref{eq:stateconst}.

Finally, the input constraints are handled similarly, that is, condition \eqref{eq:contconstr} is satisfied if and only if for all $l = 1, \ldots \mathcal N_u$,
\begin{equation}
	e_l^T \left( H_u K x - h_u \right) \leq 0
\end{equation}
for all $x \in \mathcal X$.
By the S procedure this can be translated into the positive definiteness of the matrix
\begin{equation}
	\left[\begin{smallmatrix}
		H_x^T Z^T D_u^l Z H_x & - \frac{1}{2} K^T H_u^T e_l \\ * & e_l^T h_u - \mathds{1}^T D_u^l \mathds{1}
	\end{smallmatrix}\right] \succ 0.
\end{equation}
Applying the congruency transform $\blkdiag{H_x^{-T}, I}$ gives \eqref{eq:lmi_fifth} for the same $\hat K$ as before.

We now argue that $(H_x, \hat K H_x)$ that satisfy \eqref{eq:lmi_first} and \eqref{eq:lmi_third} - \eqref{eq:lmi_fifth} are indeed a solution to Problem \ref{probl:main}. By definition, both $H_x$ and $\hat K H_x$ are block diagonal in the sizes of the subsystems, so we can extract sets $\mathcal X_i$ (from blocks of $H_x$) and local feedback controllers $K_i$ (from blocks of $\hat K H_x$) that render $\mathcal X_i$ invariant. The feedback controller provides for each $x_i \in \mathcal X_i$ a control $K_i x_i$ that enforces invariance of $\mathcal X_i$. This shows that \eqref{eq:invariancespec} is satisfied.

\IEEEtriggeratref{9}
\bibliographystyle{IEEEtran}
\bibliography{IEEEabrv,references}

\end{document}